\documentclass[journal]{IEEEtran}

\usepackage{amsmath}
\usepackage{amssymb, latexsym}
\usepackage{array}
\usepackage{graphicx}
\usepackage{cite}

\newcommand{\eqn}[1]{(\ref{#1})}
\newcommand{\fig}[1]{Fig. \ref{#1}}
\newcommand{\Fig}[1]{Figure \ref{#1}}
\newcommand{\Figs}[1]{Figures \ref{#1}}
\newcommand{\figs}[1]{Figs. \ref{#1}}

\begin{document}
\title{Phase and Amplitude Responses of Narrow-Band Optical Filter Measured by Microwave Network Analyzer}

\author{Hsi-Cheng Wang,~\IEEEmembership{Student Member,~IEEE}, Keang-Po Ho,~\IEEEmembership{Senior Member,~IEEE} and Hau-Kai Chen, and~Hsin-Chia~Lu~\IEEEmembership{Member,~IEEE}\\
\thanks{Manuscript received March 21, 2006, revised ??, 2006. %
This research was supported in part by the National Science Council of Taiwan under Grant NSC-94-2219-E-002-023 and NSC-94-2219-E-002-024.}
\thanks{H.-C. Wang and H.-K. Chen are  with the Institute of Communication Engineering, National Taiwan University, Taipei 106, Taiwan.}
\thanks{K.-P. Ho is with SiBEAM, Sunnyvale, CA 94086, on leave from the Institute of Communication Engineering and Department of Electrical Engineering, National Taiwan University, Taipei 106, Taiwan. (Tel: +1-650-305-1026, Fax: +886-2-2368-3824, E-mail: kpho@ieee.org)}
\thanks{H.-C. Lu is with the Institute of Electronics Engineering and Department of Electrical Engineering, National Taiwan University, Taipei 106, Taiwan. }
}

\markboth{Journal of Lightwave Technology}{H.-C. Wang {\em et al.}: Accurate Narrow-Band Optical Filter Characterization Using Network Analyzer}

\maketitle

\begin{abstract}
The phase and amplitude responses of a narrow-band optical filter are measured simultaneously using a microwave network analyzer.
The measurement is based on an interferometric arrangement to split light into two paths and then combine them.
In one of the two paths, a Mach-Zehnder modulator generates two tones without carrier and the narrow-band optical filter just passes through one of the tones.
The temperature and environmental variations are removed by separated phase and amplitude averaging.
The amplitude and phase responses of the optical filter are measured to the resolution and accuracy of the network analyzer.
\end{abstract}

\begin{keywords}
Narrow-band Optical Filter, Phase Measurement, Dispersion Measurement
\end{keywords}

 \section{Introduction}

\PARstart{T}{he} chromatic dispersion, or equivalently the phase response, of a narrow-band optical filter is always difficult to measure.
When the widely used modulation phase-shift method \cite{costa82, ansi} is applied for narrow-band measurement, both the accuracy and resolution are limited \cite{niemi01}.
Typical measurement equipment of the phase-shift method quotes the resolution of the tunable laser as the resolution for dispersion measurement that is actually limited by the modulation frequency.
High modulation frequency gives better phase accuracy but limits the frequency or wavelength accuracy or resolution bandwidth.
Low modulation frequency gives better frequency accuracy but limits the phase accuracy.
Dual-frequency phase-shift method may obtain better accuracy \cite{fortenberry03}.


Recently, narrow-band optical filter was used for time-resolved optical filtering (TROF) to simultaneously measure the amplitude and phase of an optical signal with \cite{ho05a} or without chromatic dispersion \cite{ho0601}.
The dual of frequency-resolved optical gating (FROG) \cite{trebino97}, TROF needs an accurate tunable narrow-band optical filter to obtain the spectrogram.
Ideally, the tunable optical filter of \cite{wildnauer93} in the measurement of \cite{ho05a, ho0601} does not have dispersion.
An accurate measurement is required to confirm the dispersionless assumption in \cite{wildnauer93, ho05a, ho0601}.

Optical filter or its equivalent is widely used in dense wavelength-division-multiplexed (DWDM) systems \cite{bigo04, rasmussen04}.
The group delay ripples of DWDM filter should be minimized for better system performance \cite{niemi01}.
Typical DWDM systems have channel spacing of 50 or 100 GHz, corresponding to a wavelength separation of 0.4 or 0.8 nm.
Future DWDM systems may potentially have narrow channel spacing using narrow-band optical filters \cite{yu03}.

Narrow-band optical filter is recently used for chirp control \cite{mahgerefteh05, yan05, chen05, matsui06}.
Those optical filters for chirp control typically have a narrower bandwidth than DWDM applications.
The phase response or dispersion for those chirp control filters should not induce further chirp into the signal.
Conventional modulation phase-shift method is not accurate for those narrow-band optical filters.

Unlike modulation phase-shift method suitable for wide-band devices like optical fiber \cite{ansi}, an interferometric method is used here to simultaneously measure the amplitude and phase responses of a narrow-band optical filter using a wide-band microwave network analyzer.
The responses of the optical filter are measured to the amplitude and frequency accuracy of the network analyzer.
When the optical filter of \cite{ho05a, ho0601, wildnauer93} is measured, the frequency resolution is down to 25 MHz and the frequency accuracy or resolution bandwidth is down to 1 kHz.
The measurement result is crucial for TROF to retrieve electric field of an optical signal.
The interferometric method here is especially suitable for a narrow-band device.

The remaining parts of this paper are organized as following: Sec.~\ref{sec:oper} explains the operation principle of the method to measure the amplitude and phase responses of a narrow-band optical filter; Sec.~\ref{sec:meas} presents the measurement results for the double-pass monochromator in \cite{wildnauer93} with an optical bandwidth of 5 GHz; and Secs.~\ref{sec:dis} and \ref{sec:end} are the discussion and conclusion, respectively.

 \section{Operation Principle of the Measurement}
\label{sec:oper}

\begin{figure}
 \centerline{\includegraphics[width = 0.45 \textwidth]{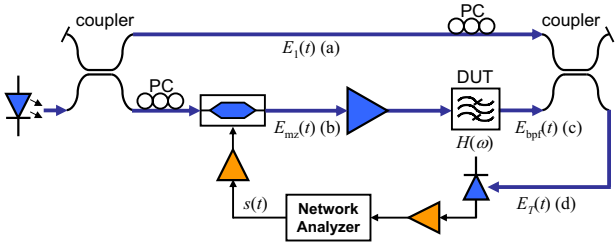}}
 \caption{System block diagram of experimental setup to measure amplitude and phase responses of a narrow-band optical filter. (DUT: device-under-test, PC: Polarization Controller)}
 \label{figsetup}
 \end{figure}

\Fig{figsetup} shows the schematic diagram for the experimental setup to characterize a narrow-band optical filter.
The setup of \fig{figsetup} is based on a Mach-Zehnder interferometric arrangement.
The optical carrier from a narrow linewidth laser source is splitted into two paths.
These two paths are combined after approximately the same path delay.

The microwave network analyzer is the major equipment in the setup of \fig{figsetup}.
To measure the transfer characteristic of a device at the microwave frequency of, for example, $\omega_m$, the network analyzer sends a sinusoidal signal at $\omega_m$ from its output port.
At the same time, the input port receives the signal after the device-under-test and the amplitude and phase of the sinusoidal signal at $\omega_m$ are measured.
Equivalently speaking, there is a very narrow-band electrical filter centered at $\omega_m$ at the receiver and followed by a precision vector voltage meter to find both amplitude and phase.
Ideally, both the noise and interference at other microwave frequencies beyond its resolution bandwidth do not affect the measurement accuracy.

The upper path of \fig{figsetup} just passes the signal to the combiner after some delay.
The lower path of \fig{figsetup} includes a high-speed Mach-Zehnder modulator (MZM) driven by the network analyzer and followed by the narrow-band optical filter under test.
The combined signal from the two paths is detected by a fast photodiode and feeds to the network analyzer.
The MZM is operated at the minimum transmission point.
The measurement operates better for a narrow-band optical filter with a bandwidth less than $\omega_m$.

\begin{figure}
\begin{center}
\begin{tabular}{cc}
\includegraphics[width = 0.225 \textwidth]{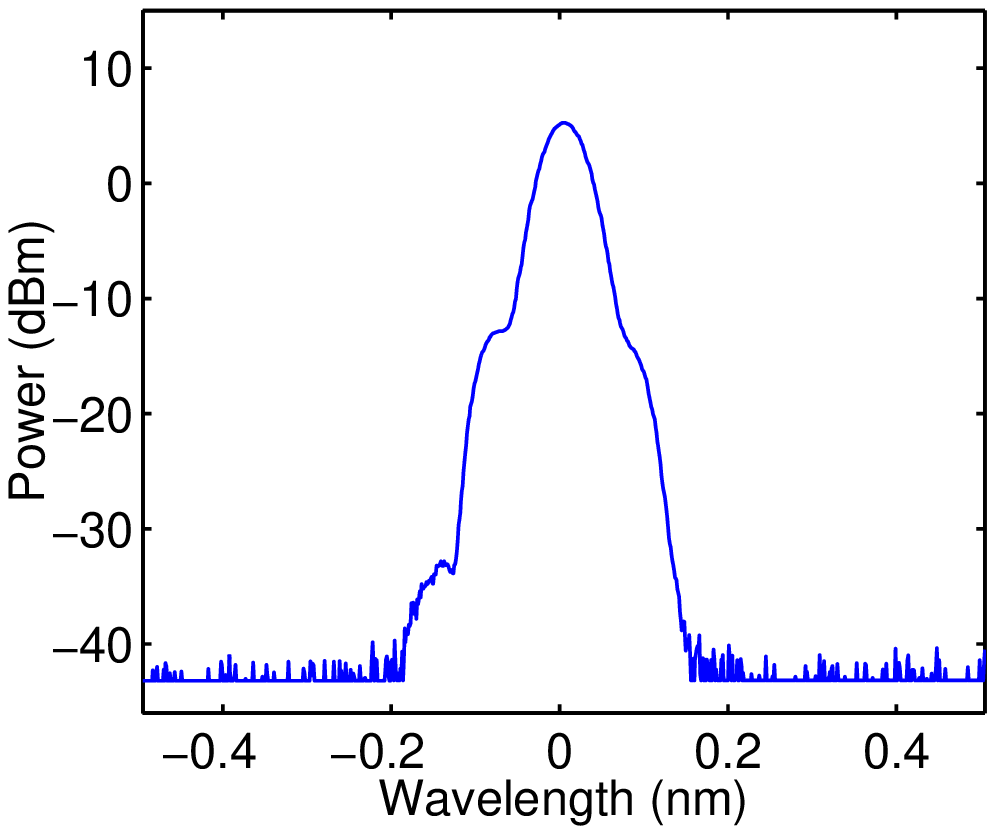} &
\includegraphics[width = 0.225 \textwidth]{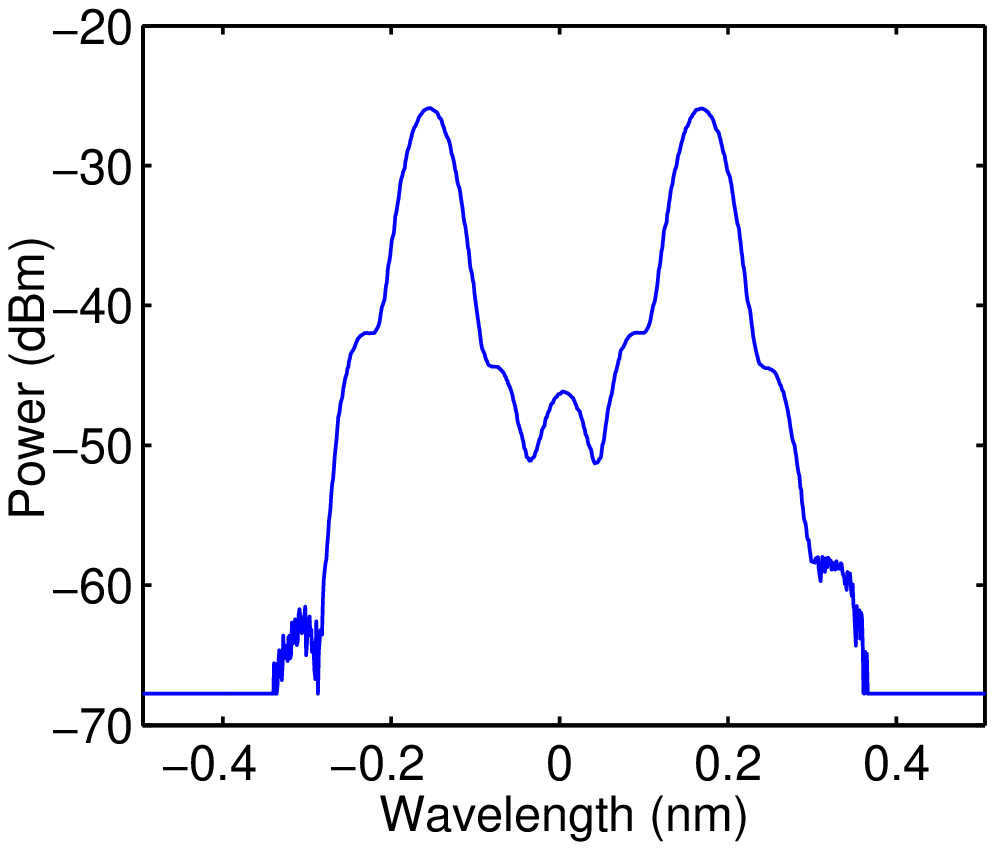} \\[-2pt]
 {\small (a) } & {\small (b)} \\[0.1cm]
\includegraphics[width = 0.225 \textwidth]{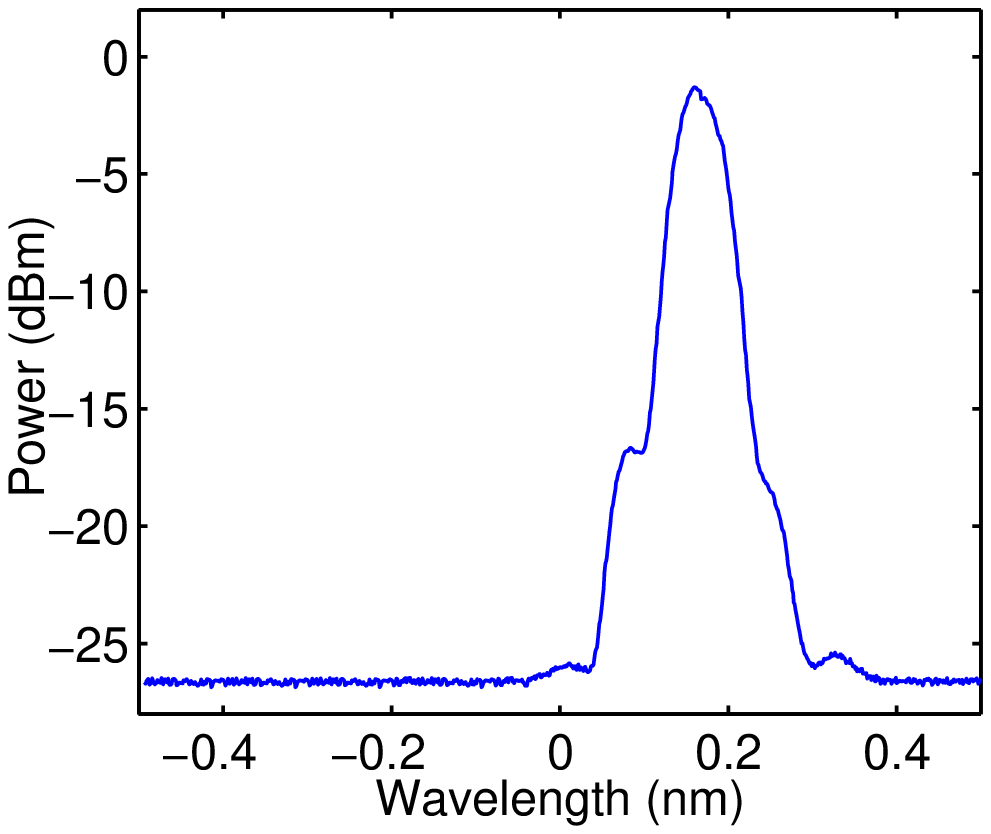} &
\includegraphics[width = 0.225 \textwidth]{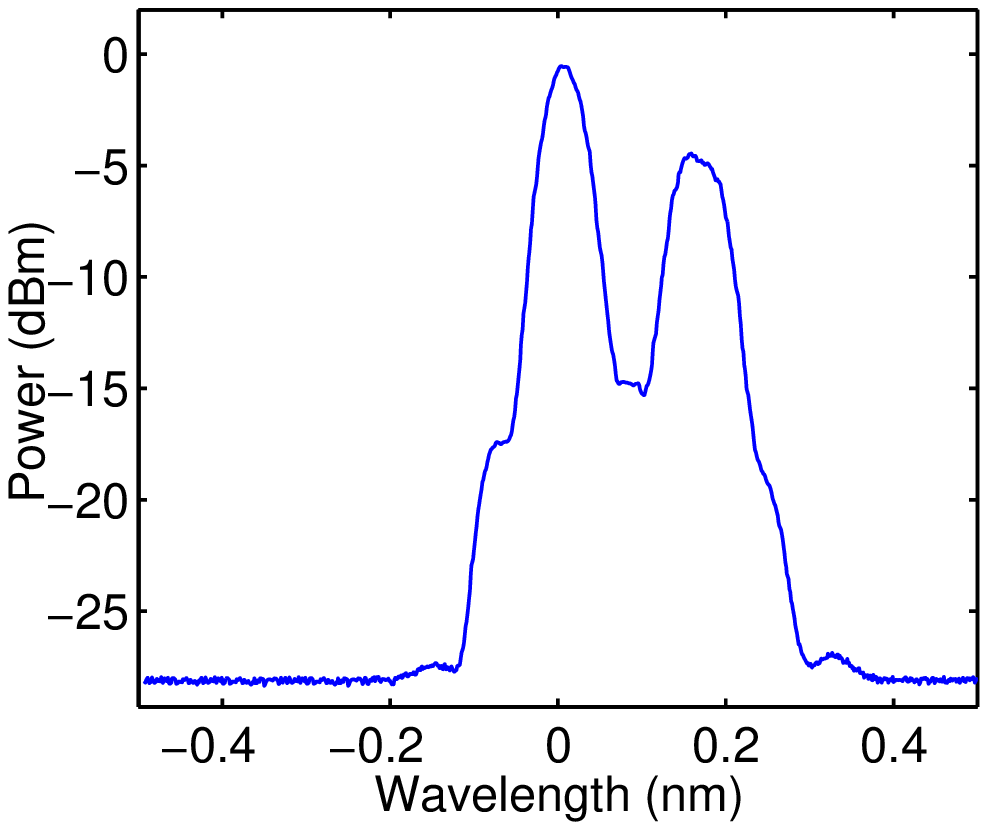} \\[-2pt]
{\small (c)}  & {\small (d)}
\end{tabular}
\end{center}
\caption{The spectrum of (a) laser source, (b) MZM output, (c) optical filter output, and (d) combiner output, where the spectra of (a) and (b) are measured
and the spectra of (c) and (d) are simulated results.}
\label{figsp}
\end{figure}

Ideally, the narrow-band optical filter has a bandwidth far smaller than $\omega_m$ and centers at, for example and without loss of generality, $\omega_c - \omega_m$, where $\omega_c$ is the angular frequency of the laser source.
Because of the narrow-band assumption, the relationship of $\left|H(\omega_c - \omega_m) \right| \gg \left|H(\omega_c + \omega_m) \right|$ is valid, where $H (\cdot )$ is the frequency response of the narrow-band optical filter.
For better illustration, \Figs{figsp} show the measured and simulated spectra with filter centered at $\omega_c - \omega_m$ and a fixed modulation frequency of $\omega_m = 20$ GHz for \fig{figsetup}.
The spectra in \fig{figsp} are for illustration purpose and the lower path actually has a frequency $\omega_m$ less than the upper path due to the inverse relationship between frequency and wavelength.

\Figs{figsp}(a) and (b) are the spectra of the laser source and output of MZM as measured by an optical spectrum analyzer, respectively.
The spectrum of \fig{figsp}(a) is a single tone at $\omega_c$ that passes directly to the combiner.
The spectrum of \fig{figsp}(b) is the output of the MZM with two tones at $\omega_c \pm \omega_m$ as the MZM is biased at the minimum transmission point.
\Figs{figsp}(c) and (d) show computer-simulated spectra for the signal passing through optical filter and after the combiner.
The spectrum of \fig{figsp}(c) shows that the optical filter just passes through one of the tones of the MZM output of \fig{figsp}(b).
The spectrum of \fig{figsp}(d) is the combination of the spectrum of \figs{figsp}(a) and (c) for the signal after the combiner.
The correct operation of the setup requires that the two tones of the combined signal of \fig{figsp}(d) is separated by $\omega_m$, the same angular frequency for the driving signal of the microwave network analyzer.

The output signal of the upper path of \fig{figsetup}, ignores some factors of constant delay and constant loss, is proportional to $E_1(t) = E_0 \exp \left(j \omega_c t \right)$, where $\omega_c$ and $E_0$ are the angular frequency and the complex envelope of the laser source.
The spectrum of the signal at upper path is the same as that from laser source shown in \fig{figsp}(a).

The drive signal to the MZM is assumed as $s(t) = A_m \sin( \omega_m t + \varphi_m)$, where $A_m$ and $\varphi_m$ are amplitude and phase of the driving signal, respectively.
Here, we assume that both $A_m$ and $\varphi_m$ are two known constants.
In practice, both $A_m$ and $\varphi_m$ depend on the frequency responses of the network analyzer output driver, modulator driver, and MZM that can be calibrated out together with some other factors.
Ignored some constant factors, the MZM with minimum bias and modulated signal $s(t)$ has output of \cite{hobook}
 \begin{equation}
 E_\mathrm{mz}(t) =E_0 \exp\left( j \omega_c t \right) \sin \left [\frac{\pi}{2} \frac{s(t)}{V_{\pi}}  \right],
 \label{modout}
 \end{equation}
where $V_\pi$ is the voltage to give a $\pi$ phase shift for the modulator. The bias condition of \eqn{modout} is widely used in  binary phase-modulated transmitter \cite{hobook, gnauck05} and frequency-doubling photonic mixer \cite{ho97}.
With minimum-biased operation, the electric field at the MZM output is
\begin{multline}
E_\mathrm{mz}(t) = 2j E_0 \exp\left( j \omega_c t \right) \\ \times \sum_{k=0}^{\infty} \! J_{2k+1} \left(\beta_m \right)\sin\left[ (2k+1)(\omega_m t + \varphi_m) \right],
\label{modbessel}
\end{multline}
\noindent where $J_n(\cdot)$ is the $n$th-order Bessel function of the first kind, and $\beta_m = \pi A_m/2 V_\pi$ is the modulation index of the signal.
The MZM output of \eqn{modbessel} is the tones at $\omega_c \pm (2 k + 1) \omega_m$ and dominates by the first-order tone of $\omega_c \pm \omega_m$.
In terms of power, the power given by the term of  $J_1(\beta_m)$ is about 30-dB larger than $J_3(\beta_m)$ for modulation index of $\beta_m$ smaller than $0.85$.
The measured spectrum of \fig{figsp}(b) just has two major tones at $\omega_c \pm \omega_m$.
Just to keep the leading term of $k=0$  for $J_1(\beta_m)$, the modulator output of \eqn{modout} is proportional to
 \begin{equation}
 E_\mathrm{mz}(t) = E_0 J_1\left(\beta_m \right)
 \left[  e^{ j (\omega_c + \omega_m) t + j \varphi_m }
         -  e^{ j (\omega_c - \omega_m) t - j \varphi_m }
 \right],
 \label{twotones}
 \end{equation}
with two equal tones at both $\omega_c \pm \omega_m$.
In practical system, as discussed later, the inclusion of higher-order terms does not change the principle of the measurement.
The higher-order terms also does not contribute to the system crosstalk if the MZM is biased exactly at the minimum transmission point.

With the two-tone input of \eqn{twotones} and with the narrow-band assumption of $|H(\omega_c - \omega_m)| \gg |H(\omega_c + \omega_m)|$ , the output of the optical filter is equal to
 \begin{equation}
 E_\mathrm{bpf}(t) = E_0  J_1\left(\beta_m \right) H( \omega_c - \omega_m)  e^{ j (\omega_c - \omega_m) t - j \varphi_m}.
 \label{bpfout}
 \end{equation}
\Fig{figsp}(c) shows the simulated spectrum of the filter output of \eqn{bpfout}.
Without taking into account many constant factors, the combiner output of \fig{figsetup} is the summation of $E_1(t) +  E_\mathrm{bpf}(t)$, or
 \begin{equation}
 E_T(t) = E_0 e^{ j \omega_c t}       \left[  1 +  J_1\left(\beta_m \right) H( \omega_c - \omega_m) e^{ -j \omega_m t - j \varphi_m }  \right],
 \label{comb}
 \end{equation}
with two tones at $\omega_c$ and $\omega_c - \omega_m$.
These two tones with spectrum of \fig{figsp}(d) have a frequency difference of $\omega_m$ that is exactly the same as the operating frequency of the network analyzer.

The high-speed photodiode gives a photocurrent proportional to $\left| E_1(t) +  E_\mathrm{bpf}(t) \right|^2$.
As the network analyzer just processes the signal around $\omega_m$, the received signal of the analyzer is beating of \eqn{comb} having a frequency exactly the same as $\omega_m$ of
\begin{equation}
s_{\mathrm{na}}(t)
 = 2 E_0^2 J_1\left(\beta_m \right) \Re \left\{ D_m H( \omega_c - \omega_m)
     e^{ -j \omega_m t - j \varphi_m} \right\},
\label{sna}
\end{equation}
where $D_m$ is the complex frequency response of the combination of the photodiode, the signal amplifier afterward, and the network analyzer receiver, and $\Re\{ \cdot \}$ denotes the real part of a complex number.
Ignored factors independent of frequency, the network analyzer measures a complex response of
\begin{equation}
J_1\left(\beta_m \right) D_m H( \omega_c - \omega_m)  e^{-j \varphi_m}.
\label{naresp}
\end{equation}
In the derivation of \eqn{naresp} from \eqn{modout} to \eqn{sna}, many constant factors are ignored by replacing proportional by equality relationship.
Because those factors are independent of frequency, the inclusion or exclusion of those factors does not change the measured response of \eqn{naresp} but just moves the whole curve up and down.

Without the optical filter, followed the derivation from \eqn{modout} to \eqn{naresp}, the network analyzer gives a complex response proportional to $J_1\left(\beta_m \right)  D_m  e^{- j \varphi_m}$.
The measurement with the same setup but without the optical filter gives the calibration factors.
Excluding the calibration factors from the measured response of \eqn{naresp} gives the optical filter response of
\begin{equation}
H( \omega_c - \omega_m).
\end{equation}

In practical measurement, the path length of the interferometric structure of \fig{figsetup} changes slightly with time due to temperature and other environmental variations.
The phase response changes slowly even within the same frequency scanning.
With the random phase due to temperature and environmental variations, the frequency response of \eqn{naresp} becomes
\begin{equation}
J_1\left(\beta_m \right)  D_m H( \omega_c - \omega_m)
     e^{ - j \varphi_m + j\Theta(t)} ,
 \label{sna1}
\end{equation}
where the $\Theta(t)$ is the random phase contributed from the phase variation of two paths.
As the default setting for typical network analyzer, if the complex number of \eqn{sna1} is averaged over time, the averaged complex response is equal to zero.
Because the environmental phase of $\Theta(t)$ changes slowly with time and is zero mean, as shown numerically later, the average of many responses of \eqn{sna1} separately in both amplitude and phase take out the random phase of $\Theta(t)$.
Unfortunately, separated amplitude and phase averaging is not a function provided by typical network analyzer.
In phase averaging, the whole phase curve may move up and down but the profile or shape of the curve does not change with time.

\section{Measurement Results}
\label{sec:meas}

The practical measurement needs to take into account the optical power budget, signal-to-noise ratio of the microwave signal, length matching of the upper and lower paths, signal-to-crosstalk ratio, and measurement calibration.
The measurement setup here follows the schematic diagram of \fig{figsetup}.
With a fixed wavelength of 1554 nm, Agilent 81640A tunable laser is used as a fixed wavelength laser for its low linewidth and high power of 6 dBm.
The signal is splitted and combined by two 3-dB couplers.

The signal output at a 50-GHz network analyzer (Agilent E8364B) is 0 dBm and boosted up to 17 dBm ($A_m \approx 2.24$ V) by a broadband amplifier (SHF-824) to drive a 40-GHz zero-chirp MZM (Avanex SD-40) with $V_{\pi} = 5.5$ V.
The modulation index of $\beta_m$ is about  $0.7$.
In terms of power, the third-order tone of $J_3(\beta_m)$ is about -33 dB of the first-order tone.
The MZM output is amplified by an Erbium-doped fiber amplifier (EDFA) with small-signal gain of 20 dB at 1554 nm and saturation output power of 22 dBm.
The narrow-band optical filter is the monochromator of Agilent 86146B with full-wave-half-maximum (FWHM) bandwidth of 0.04 nm (5 GHz) and insertion loss about 10 dB \cite{ho05a, wildnauer93, chen05}.
Because of the loss of both MZM and the monochromator, the lower path of \fig{figsetup} with the narrow-band optical filter typically has a power about $10$ dB less than the upper path at the combiner.

Both the upper and lower paths have a length of about $48 \pm 0.5$ m to maintain a clean phase response for the measurement.
A polarization controller is used in the upper path to align the polarization of the two paths.
To improve the network analyzer sensitivity, an additional broadband amplifier (SHF-803) is used after a 50-GHz photodiode (u$^2$t XPDV2020R).
The calibration measurement is provided when the Agilent 86146B monochromator is tuned to a bandwidth of 10 nm, equivalently without an optical filter.

\begin{figure}
 \centerline{\includegraphics[width = 0.5 \textwidth]{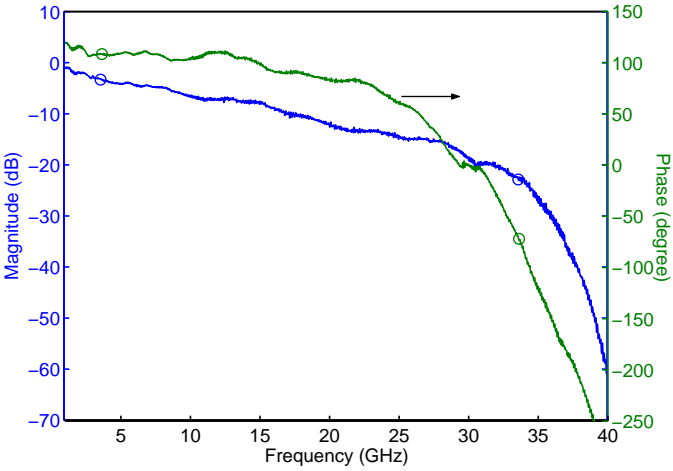}}
 \caption{The calibration measurement of amplitude and phase responses with circle marks denoting the 30-GHz band to characterize the  narrow-band optical filter.}
 \label{figcal}
 \end{figure}

With an optical filter bandwidth of 10 nm, \Fig{figcal} shows the measured amplitude and phase frequency response for calibration purpose.
The center frequency of the filter for calibration is 0.145 nm (18.6 GHz) from the carrier frequency.
The calibration traces shows the composite frequency response of the broadband amplifiers, MZM, and photodetector, where the filter response is flat within the 30-GHz band for a 10-nm bandwidth optical filter, corresponding to a frequency bandwidth of about 1250 GHz.
The calibration trace is averaged over 64 measured traces to take out the environmental variations.
The four circles of \fig{figcal} shows the beginning and end of the 30-GHz band that is used to characterize the narrow-band optical filter.

\begin{figure}
 \centerline{\includegraphics[width = 0.5 \textwidth]{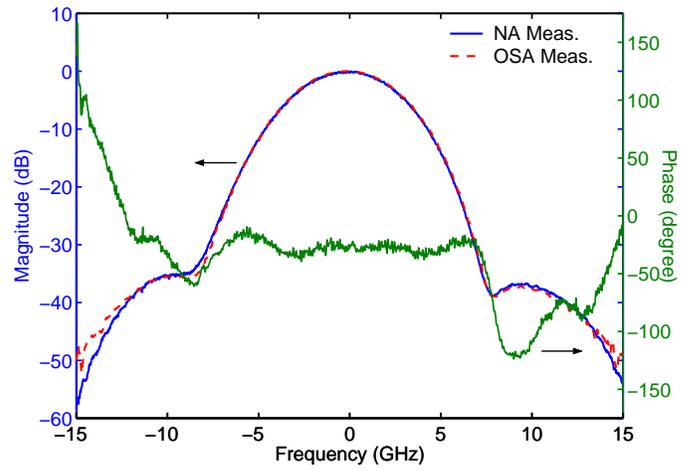}}
 \caption{Amplitude and phase responses of the narrow-band optical filter measured by the setup of \fig{figsetup}.
The dashed-line is the amplitude measured in both \cite{ho05a, ho0601}.}
 \label{figtr}
 \end{figure}

\Fig{figtr} shows the measured amplitude and phase frequency responses of the narrow-band optical filter after calibration.
\Fig{figtr} also shows the amplitude response measured in both \cite{ho0601, ho05a}.
The center wavelength of the optical filter is 0.145 nm (18.6 GHz) from the carrier frequency, the same as that for \Fig{figcal}.
\Fig{figtr} is shifted by 18.6 GHz to center the optical filter to zero frequency.
In order to avoid low signal-to-crosstalk band, the \fig{figtr} shows the combination of the filter measurement located at either side tone ($\omega_c \pm \omega_m$) with high signal-to-crosstalk part.
The original data are measured by network analyzer from 1 to 40 GHz with frequency step of 25 MHz and a resolution bandwidth of 1 kHz.
The network analyzer measured response in \fig{figtr} is averaged over 64 measured data.
Better resolution is available from the network analyzer but requires longer measurement time.
The measurement of \fig{figtr} shows that the phase frequency of the optical filter is very flat or more precisely $\pm 12^{\circ}$ phase change in the important region of $\pm7.5$ GHz, confirming the assumption of \cite{ho05a, ho0601} that the monochromator does not have dispersion.

\section{Discussion}
\label{sec:dis}

The complex response of \eqn{naresp} includes the response of $H(\omega_c - \omega_m)$ from the narrow-band optical filter.
The measurement setup of \fig{figsetup} uses an MZM to generate two optical tones at $\omega_c \pm \omega_m$, respectively.
Without loss of generality, the narrow-band optical filter just selects one of the tones at $\omega_c - \omega_m$.
Both the amplitude and phase responses of the optical filter are also embedded in the output signal at the frequency of $\omega_c - \omega_m$.
When this signal beats with the original carrier signal at $\omega_c$, the filter response of $H(\omega_c - \omega_m)$ is measured by the microwave network analyzer.
The correct operation of the setup of \fig{figsetup} requires the condition that $|H(\omega_c - \omega_m)| \gg |H(\omega_c + \omega_m)|$.
This condition can translate to the requirement that the operating range of the setup of \fig{figsetup} is about four to five times larger than the bandwidth of the optical filter.

The measurement setup can measure an optical filter with wider bandwidth if the center frequency of $\omega_c$ is tuned such that only the large modulated frequency of $\omega_m$ is used.
Like \fig{figtr}, the frequency response is combined from the measurements when $\omega_c$ is tuned to either smaller or larger than the center of the optical filter.
With this modified measurement procedure, the bandwidth of the optical filter can be in the range of about 10 to 15 GHz.
Used for demonstration purpose, this special technique is not required for the measurement of the filter of \fig{figtr}.

The optical filter with a bandwidth of 0.04 and 10 nm has a loss difference of about 2 dB.
The difference in the filter loss does not change the filter profile.
When the calibration trace is obtained by tuning the bandwidth of the monochromator to 10 nm, the measured amplitude and phase responses are contaminated by a larger amount EDFA noise.
The effect of EDFA noise is reduced by averaging over many measured traces.
The calibration trace of \fig{figcal} is obtained by averaging of 64 traces.

>From the measurements in \fig{figtr}, the amplitude response measured by network analyzer is matched very well to that measured in \cite{ho05a}.
In the measurement for \cite{ho05a}, a single-wavelength laser source inputs to the monochromator followed by an optical power meter.
The monochromator is tuned to slightly different wavelength to obtain the amplitude response profile.
The amplitude response is derived from the output power as a function of the center wavelength of the monochromator.
The variation of the phase response of \fig{figtr} is also very small as explained by the physics of the double-pass monochromator in \cite{wildnauer93}.
The non-symmetric magnitude response is caused by manufacturing error of the device and agrees with the measurement in \cite{lewis88}.

In the measurement setup of \fig{figsetup}, the bias point of the MZM is chosen by minimizing the power at the carrier frequency at the MZM spectrum output.
The MZM should be minimum bias by this tuning method.
Ideally, there is no carrier tone at the frequency of $\omega_c$.
In the measurement of \fig{figsp}(b), the tone at $\omega_c$ is about 21.5 dB smaller than the two tones at both $\omega_c \pm \omega_m$ due to the limited extinction ratio of the MZM.
As from \fig{figsp}(c), the tone at the carrier frequency of $\omega_c$ can be ignored as the response of $H(\omega_c)$ is far smaller than the response at $H(\omega_c - \omega_m)$.
At the input of the combiner, the upper path of \fig{figsetup} gives an optical signal 10 dB larger than that of the lower path.
In additional to the difference in $H(\omega_c)$ and $H(\omega_c - \omega_m)$, the crosstalk is at most 30 dB lower than the signal power.
As from \eqn{modbessel}, the higher-order term does not give crosstalk to the signal.
The only crosstalk is from the non-zero carrier at $\omega_c$ and the insufficient difference in $H(\omega_c \pm \omega_m)$.
Even in the worst case, the crosstalk is 40 to 50 dB lower than the signal.

The minimum bias operation of an MZM has larger insertion loss than that of maximum bias.
Without driving signal from network analyzer, the minimum- and maximum-biased MZM has about 40 dB and 4 dB insertion loss, respectively.
The extinction ratio of the MZM is about 36 dB from this measurement.
Although the maximum-biased MZM provides low insertion loss, the output signal has a large carrier tone that is not suitable for this measurement propose.
With driving signal, the minimum-biased MZM has about 30 dB insertion loss by comparing input power with averaged output power.
The power budget is critical to the setup with MZM that operates at minimum bias point.

The network analyzer provides dynamic range of up to 110 dB and noise floor is down to about -90 dBm over 1 to 40 GHz band with resolution bandwidth of 1 kHz.
While the narrow-band filter is centered at 18.6 GHz from carrier, the maximum power of the amplitude is about -17 dBm and the range of magnitude response is bounded within 75 dB over 1- to 40-GHz band.

In the measurement setup, the length of lower path is measured by its equivalent length of a single mode fiber.
The delay of the lower path is about 236 ns with equivalent length of 47.7 m, where delay of a 10 m single mode patch cord is measured to be 49.4 ns.
The upper path has a 45 m patch cord together with a 3 m polarization controller.

The phase response of the measurement hurts by environmental instability.
When path difference between upper and lower path is reduced down to 0.5 m, temperature variation of two paths still give phase variation.
The longest delay is due to EDFA with about 38 m for the lower path and compensates by the patch core for upper path.
The temperature induced length variation converts to phase variation that cannot be averaged out by the network analyzer itself.
By unwrapping the phase response measured by the network analyzer, the amplitude and phase responses can be averaged separately.

There are 1601 resolution points in each measured trace over 1- to 40-GHz band. Requiring additional delay for each resolution points, we choose
stepped sweep mode for the network analyzer to reduce the measurement error due to the long path delay of the setup. The measurement acquisition
time is about 2 second/trace and about 1000 traces are collected for detail data analysis.
The measured and calibration traces are averaged over 64 traces to
mitigate the amplifier noise from EDFA and environmental variations.
With 64 trace moving averaging, the measurements are very stable with time.

 \begin{figure}
 \centerline{\includegraphics[width = 0.4 \textwidth]{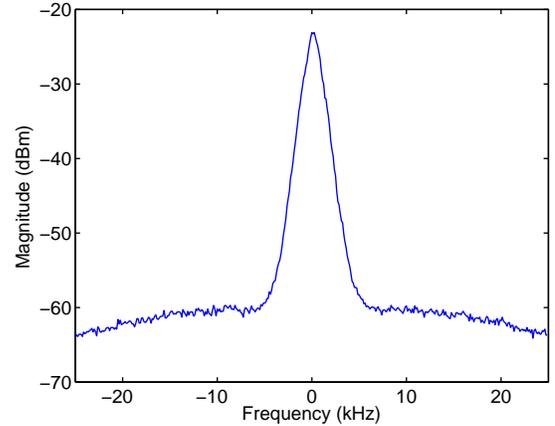}}
 \caption{The linewidth of the laser source measured by microwave spectrum analyzer centered at 20 GHz.}
 \label{figlw}
 \end{figure}

\Fig{figtr} is averaged over 64 traces.
The phase of \fig{figtr} at zero frequency is determined by the first trace.
The phase of other 63 traces at zero frequency is moved to the specific value after the phase unwrapping.
Consider the 64-moving averaging of 1000 traces and within $\pm 7.5$ GHz of \fig{figtr}, there is less than  $\pm 3^{\circ}$ of peak phase variations.
For low signal power region, the 64-averaged traces has phase variation of $\pm 10^{\circ}$ of right- and
left-most edge of the measured phase response due to additional thermal noise. The thermal effect becomes significant for the network analyzer
when received signal power is down to -70 dBm or less, where the signal power at two edges in \fig{figtr} are about -85 dBm. For smaller average
window size of 16 and 32 sliding over 1000 measured traces, the peak phase variations are within $\pm 10^{\circ}$ and $\pm 6^{\circ}$ in $\pm
7.5$ GHz measurement band, respectively. The amplitude responses after average of 16, 32, and 64 traces are very stable with indistinguishable
variation in important region of $\pm 7.5$ GHz. The 64-averaged result shows its good stability over 1000 traces or equivalently 50 mins
measurement time.

For commercial instrument (for example, Agilent 86038B) using traditional modulation phase-shift method, the frequency and wavelength step size (or resolution) is down to 12.5 MHz (0.1 pm) in accordance with the finest step of the tunable laser.
The resolution bandwidth for dispersion of the equipment is determined by the modulation frequency.
The typical absolute and relative wavelength accuracy are $\pm 3.6$ pm (450 MHz) and $\pm 2$ pm (250 MHz).
The better wavelength accuracy using smaller modulation frequency gives lower phase accuracy.
The accuracy for phase-shift method is insufficient for narrow-band device.
As an example, the Agilent 86038B has the smallest modulation frequency of 5 MHz with a phase (or dispersion) accuracy 400 times worse than the largest modulation frequency of 2 GHz.
Assume a dispersion coefficient of $D = 17$ ps/km/nm and a measurement range of $\pm1$ GHz, the phase-shift method must measure a phase change within $\pm 0.00025^\circ$ for modulation frequency of 5 MHz but a phase change within $\pm 0.1^\circ$ for modulation frequency of 2 GHz.
Even for the modulation frequency of 2 GHz and using an intermediate frequency bandwidth of 70 Hz, a phase change within $\pm 0.1^\circ$ is not simple.

The frequency accuracy of the network analyzer is about 1 ppm or less than 50 kHz for a frequency up to 50 GHz.
The resolution bandwidth for network analyzer can be down to the order of 1 to 10 Hz.
With a step size of 25 MHz, the measurement for \fig{figtr} uses a resolution bandwidth of 1 kHz.
Although the step size for tunable laser in phase-shift method is comparable to that in network analyzer, the accuracy or resolution bandwidth for phase-shift method is many orders larger than that for network analyzer method proposed here.
\Fig{figlw} shows the beating of the combined signal of \eqn{sna} as measured by a spectrum analyzer when the modulation frequency of $\omega_m$ is fixed at 20 GHz with a resolution bandwidth of 1 kHz, the same as the measurement for \fig{figtr}.
The signal-to-noise ratio is about 40 dB from \fig{figlw} and the linewidth is approximately the same as the resolution bandwidth of 1 kHz.
With the signal-to-noise ratio of 40 dB, the amplitude error is less than 0.1 dB and the phase error is less than $1^\circ$ \cite[Fig. 8]{agilent1287-2}.
Both the amplitude and phase accuracy of network analyzer is better than the modulation phase-shift method.

 \section{Conclusion}
\label{sec:end}

The amplitude and phase responses of a narrow-band optical filter are measured accurately using a microwave network analyzer.
A minimum-biased MZM generates two tones at $\omega_c \pm \omega_m$ and the narrow-band optical filter selects one of the tones.
Either tone of $\omega_c \pm \omega_m$ beats with the carrier frequency of $\omega_c$ and the beating signal is sent to a network analyzer for the simultaneous measurement of amplitude and phase.

The higher-order terms of the MZM output do not degrade the measurement.
With ensemble averaging over 64 measured traces, the measurement removes temperature and environmental variations induced phase variations and EDFA noises.
Although the measurement range of the setup is limited by bandwidths of the network analyzer and the associated microwave and electro-optical components, a fine frequency resolution is provided by the network analyzer.
The measurement setup here has a step-size or frequency resolution of 25 MHz and resolution bandwidth of 1 kHz.
The frequency accuracy is determined mainly by the resolution bandwidth of the microwave network analyzer.

\section*{Acknowledgment}
The authors thank professor Joseph M. Kahn of Stanford University for very helpful suggestions on measurement issues.

\end{document}